\documentclass[twocolumn,10pt,showpacs,amsmath,lang]{revtex4}

\usepackage{longtable,array}
\usepackage{graphicx}

\usepackage{bm}

\begin{document}

\newcommand{\half}{\mbox{$\textstyle \frac{1}{2}$}}
\newcommand{\ket}[1]{\left | \, #1 \right \rangle}
\newcommand{\bra}[1]{\left \langle #1 \, \right |}
\newcommand{\beq}{\begin{equation}}
\newcommand{\eeq}{\end{equation}}
\newcommand{\bea}{\begin{eqnarray}}
\newcommand{\eea}{\end{eqnarray}}
\newcommand{\req}[1]{Eq.\ (\ref{#1})}
\newcommand{\gcc}{{\rm~g\,cm}^{-3}}
\newcommand{\Compton}{\lambda\hspace{-.44em}\raisebox{.6ex}{\mbox{-$\!$-}}%
\raisebox{-.3ex}{}_{\hspace{-1pt}{\mbox{$_\mathrm{C}$}}}}
\newcommand{\kB}{k_\mathrm{B}}
\newcommand{\omc}{\omega_\mathrm{c}}
\newcommand{\omg}{\omega_\mathrm{g}}
\newcommand{\mel}{m_e}
\newcommand{\xr}{x_{\rm r}}
\newcommand{\EF}{\epsilon_{\rm F}}
\newcommand{\Ne}{{\cal N}_B(\epsilon)}
\newcommand{\Necl}{{\cal N}_0(\epsilon)}
\newcommand{\dfde}{{\partial f^{(0)} \over\partial\epsilon}}
\newcommand{\am}{a_\mathrm{m}}
\newcommand{\dd}{{\rm\,d}}
\newcommand{\vB}{\bm{B}}
\newcommand{\dotZ}{\mbox{$\dot{\mbox{Z}}$}}
\newcommand{\msun}{\mbox{$M_\odot$}}

\title{A simple analytic model for astrophysical S-factors}

\author{D. G. Yakovlev}

\affiliation{Ioffe Physical Technical Institute, Poliekhnicheskaya
26, 194021 St.-Petersburg, Russia}

\author{M. Beard}

\affiliation{Department of Physics $\&$ The Joint Institute for
Nuclear Astrophysics, University of Notre Dame,  Notre Dame, IN
46556 USA}

\author{L. R. Gasques}

\affiliation{Laborat\'{o}rio Pelletron, Instituto de F\'{\i}sica da
Universidade de S\~ao Paulo, 05315-970, S\~ao Paulo, SP, Brazil}

\author{M. Wiescher}

\affiliation{Department of Physics $\&$ The Joint Institute for
Nuclear Astrophysics, University of Notre Dame,  Notre Dame, IN
46556 USA {} }

\date{\today}
\begin{abstract}
We propose a physically transparent analytic model of astrophysical
$S$-factors as a function of a center-of-mass energy $E$ of
colliding nuclei (below and above the Coulomb barrier) for
non-resonant fusion reactions.
For any given reaction, the $S(E)$-model contains four parameters
[two of which approximate the barrier potential, $U(r)$]. They are
easily interpolated along many reactions involving isotopes of the
same elements; they give accurate practical expressions for $S(E)$
with only several input parameters for many reactions. The model
reproduces the suppression of $S(E)$ at low energies (of
astrophysical importance) due to the shape of the low-$r$ wing of
$U(r)$. The model can be used to reconstruct $U(r)$ from computed or
measured $S(E)$. For illustration, we parameterize our recent
calculations of $S(E)$ (using the S\~ao Paulo potential and the
barrier penetration formalism) for 946 reactions involving stable
and unstable isotopes of C, O, Ne, and Mg (with 9 parameters for all
reactions involving many isotopes of the same elements, e.g., C+O).
In addition, we analyze astrophysically important $^{12}$C+$^{12}$C
reaction, compare theoretical models with experimental data, and
discuss the problem of interpolating reliably known $S(E)$ values to
low energies ($E \lesssim 2-3$ MeV).
\end{abstract}


\maketitle

\section{Introduction}
\label{s:introduct}

Nuclear reactions are very important \cite{bbfh57,fh64,clayton83}
for the structure, evolution, nucleosynthesis and various
observational manifestations of main-sequence stars, giants and
supergiants, presupernovae, white dwarfs and neutron stars.
Depending on temperature, density and other parameters, stellar
burning may involve many reactions of different nuclei, from light
to heavy, and from stable to neutron- and proton-rich ones. Their
rates can be calculated using the reaction cross sections
$\sigma(E)$, or related astrophysical $S$-factors defined as
\begin{equation}
   \sigma(E) = E^{-1}\, \exp(-2 \pi \eta)\,S(E).
\label{e:sigma}
\end{equation}
Here, $E$ is the center-of-mass energy of the reactants [$(A_1,Z_1)$
and $(A_2,Z_2)$], $\eta= \alpha/(\hbar v)=\sqrt{E_R/E}$ is the
Sommerfeld parameter, $v=\sqrt{2E/ \mu}$ is the relative velocity of
the reactants at large separations, $\alpha=Z_1 Z_2 e^2$,
$E_R=\alpha^2 \mu/(2 \hbar^2)$ is similar to the Rydberg energy in
atomic physics, and $\mu$ is the reduced mass. The factor $\exp(-2
\pi \eta)$ is proportional to the probability of penetration through
the Coulomb barrier $U(r)=\alpha/r$ with zero angular orbital
momentum, assuming that this pure Coulomb barrier extends to $r \to
0$ (for point-like nuclei); $E^{-1}$ factorizes out the well-known
pre-exponential low-energy dependence of $\sigma(E)$. The advantage
of this approach is that $S(E)$ is a much more slowly varying
function of $E$ than $\exp(-2 \pi \eta)$ and $\sigma(E)$.

For astrophysical applications, one needs to know $S(E)$ for many
reactions at low energies, $E \lesssim $ a few MeV. Experimental
measurements of $\sigma(E)$ at such energies are mainly not
available (because the Coulomb barrier exponentially suppresses
low-energy cross sections). Theoretical calculations are model
dependent, so that nuclear-physics uncertainties of calculated
$S(E)$ can be substantial. Theoretical calculations show that $S(E)$
can vary by several orders of magnitude in the energy range of
astrophysical importance for a given reaction, and it can vary over
many orders of magnitude for different reactions (e.g.,
\cite{yak2006,paper1} and references therein). It is the aim of this
paper to propose (Sec.~\ref{s:model}) a physically transparent
analytic model of $S(E)$ for non-resonant reactions between heavy
nuclei in order to explain these features, simplify the use of
available $S(E)$-data, and clarify the problem of interpolating
reliably known $S(E)$ values to low energies of astrophysical
importance.

Astrophysical $S$-factors have been parameterized by different
analytic formulae (see \cite{fh64,PWZ1969,Spillane2007,paper1}, for
references). We think that our new model is more flexible. It allows
one to approximate $S(E)$ for many reactions with minimum number of
fit parameters. For instance, our recent approximation of $S(E)$ for
946 reactions involving different isotopes of C, O, Ne, and Mg with
8514 fit parameters is now replaced (Sec.~\ref{s:example}) with the
approximation containing 90 parameters. Moreover, our model directly
relates $S(E)$ with the parameters of the effective potential $U(r)$
of nucleus-nucleus interaction, and helps to reconstruct (constrain)
$U(r)$ from the $S(E)$ data (computed or experimental ones; see
Secs.\ \ref{s:example} and \ref{s:12C}).

\section{Analytic model}
\label{s:model}


\subsection{General approach}
\label{s:general}


Let us construct a simplified model of $S(E)$ at sufficiently low
energies $E$ at which the main contribution to the reaction cross
section comes from the $s$-wave channel. For the reactions involving
the nuclei with $Z_1,Z_2 \sim 6-12$, this is true at the energies of
a few tens of MeV. According to the theory of inelastic scattering
(e.g., Ref.\ \cite{LLQM}), a reaction cross-section at such low
energies has different energy dependence above and below the Coulomb
barrier. Below the barrier (at $E < E_C$, $E_C$ being the barrier
height) it behaves as
\begin{eqnarray}
    \sigma(E)&=&S_0 E^{-1} \exp \Phi(E),
\label{e:belowsigma} \\
     \Phi(E)&= & -{2 \over \hbar}
       \int_{r_1}^{r_2} {\rm d}r\, \sqrt{2 \mu ( U-E)},
\label{e:Phi}
\end{eqnarray}
where $\Phi(E)$ is the semi-classical exponent argument in the
expression for the barrier penetrability. We adopt the
semi-classical approximation to calculate this penetrability;
$U=U(r)$ is the effective nucleus-nucleus potential (it is Coulombic
at large separations $r$ but is affected by nuclear forces at small
$r$); $r_1$ and $r_2$ are classical turning points. In
Eq.~(\ref{e:belowsigma}), $S_0$ is a slowly varying function of $E$
which we treat as a constant. Its order-of-magnitude estimate in
terms of physical quantities can be deduced, for instance, from the
consideration in Appendix C of Ref.~\cite{fh64}:
\begin{equation}
    S_0 \sim {2 \pi \hbar^2 \over \mu} \, \sqrt{E_C \over V_0},
\label{e:S0estimate}
\end{equation}
where $V_0 \sim 40$ MeV. The pre-exponent factor $E^{-1}$ in
(\ref{e:belowsigma}) can be written as $(1/\sqrt{E}) (1/\sqrt{E})$,
where one factor $1/\sqrt{E}$ is a generic feature of low-energy
reaction cross sections (neglecting barrier penetration); the extra
factor $1/\sqrt{E}$ comes from three-dimensional penetrability
through the Coulomb barrier \cite{LLQM}. Let us stress that $S_0$ is
not identical to the astrophysical $S(E)$-factor. We will see that
$S_0$, in contrast to $S(E)$, weakly depends on specific reaction.

At $E>E_C$ the Coulomb barrier is transparent (in the semi-classical
approximation), $\Phi(E)=0$, and $\sigma(E) \propto 1/\sqrt{E}$.

Combining the definition of $S(E)$, given by Eq.\ (\ref{e:sigma}),
with Eq.\ (\ref{e:belowsigma}), and using the above arguments, we
present $S(E)$ in the form
\begin{eqnarray}
   S(E)&=& S_0\,\exp \Psi(E) ,
\nonumber \\
&& \Psi(E)=2 \pi \eta + \Phi(E)
       \quad \mathrm{at~~~}E\leq E_C,
\label{e:below}\\
   S(E)&=& S_0\,\exp\left( 2 \pi \eta \right)\,\sqrt{E/E_C}
\nonumber \\
   && \times [1+\xi (E-E_C)/E]
       \quad \mathrm{at~~~}E>E_C.
\label{e:above}
\end{eqnarray}
The last equation is phenomenological and contains a constant
parameter $\xi$. This equation extends Eq.~(\ref{e:below}) to the
energies above the barrier. At energies $E \gg E_C$ (but still low
enough for the model to be valid) we have $S(E)=S_0 \exp(2 \pi \eta)
(\xi +1) \sqrt{E/E_C}$. Thus, $\xi$ determines the magnitude of the
reaction cross section at $E \gg E_C$; it is also important for
describing the $S(E)$ behavior at $E \approx E_C$.

\subsection{Model of barrier potential}
\label{s:Coulomb}


\begin{figure}[t]
\begin{center}
\includegraphics[width=8.0cm,angle=0, bb=30 170 470 590]{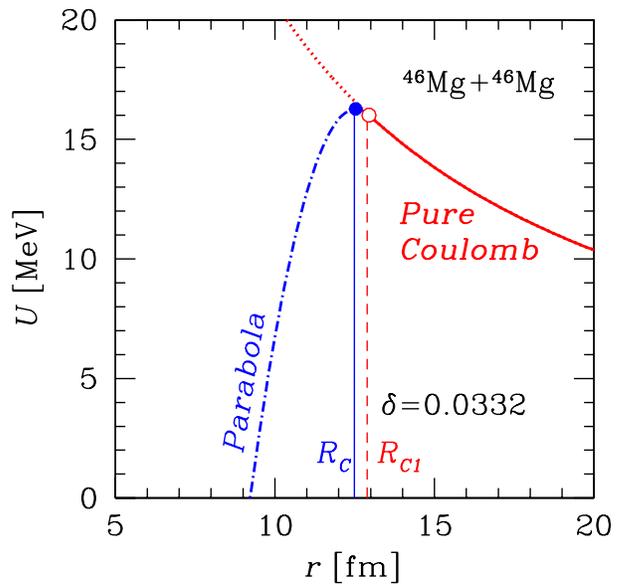}
\caption{(Color online) Model effective potential $U(r)$ for the
$^{46}$Mg+$^{46}$Mg reaction discussed in Sec.\ \ref{s:example}. The
solid curve is the pure Coulomb segment; the dot-and-dashed curve is
the inverted parabolic segment; the dotted line shows the pure
Coulomb potential extended to $r \to 0$. The filled dot marks the
barrier hight ($U=E_C$ at $r=R_C$), and the open dot separates the
Coulomb and parabolic segments ($U=E_{C1}$ at $r=R_{C1}$).}
\label{fig:Ueff}
\end{center}
\end{figure}

Let us adopt a simple model of $U(r)$,
\begin{eqnarray}
   U(r)&=&{\alpha \over r}~~~\mathrm{at}~~r\ge R_{C1},\nonumber \\
   U(r)&=&E_C \left[ 1 - \beta {(r-R_C)^2 \over R_C^2}
   \right]~~~\mathrm{at}~~r<R_{C1}.
\label{e:U}
\end{eqnarray}
It is a pure Coulomb potential at $r \ge R_{C1}$ and an inverse
parabolic potential at smaller $r$. The parabolic segment truncates
the effective interaction at small separations; $E_C=U(R_C)$ is the
maximum of $U(r)$  (the barrier height). This model is the most
natural and simple approximation of $U(r)$ which allows us to
analytically calculate the barrier penetrability. We require $U(r)$
and its derivative be continuous at $r=R_{C1}$, and introduce
$\delta=(R_{C1}-R_C)/R_C$ that characterizes the width of the peak
maximum of $U(r)$.

In this way, $U(r)$ is determined by two parameters, say, $E_C$ and
$\delta$, with
\begin{eqnarray}
  &&  R_C={ \alpha (2+3 \delta) \over 2 E_C (1+\delta)^2}, \quad
    \beta= {1 \over \delta (2+3 \delta)},
\nonumber \\
  &&  R_{C1}=R_C\,(1+\delta),
\nonumber \\
 &&   E_{C1}=U(R_{C1})=E_C\, {2+ 2 \delta \over 2 +3 \delta}.
\label{e:Upar}
\end{eqnarray}
The potential $U(r)$ passes through zero at $r=R_{C0}=R_C\, (1 -
\beta^{-1/2})$. Its behavior at smaller $r$ does not affect directly
our results. Realistic models should correspond to $\beta \gg 1$
(the low-$r$ slope of $U(r)$ should be sharp; $R_{C0}$ should be
positive) which translates into $\delta \ll {1 \over 3}$ (because
$\beta=1$ corresponds to $\delta={1 \over 3}$).

For example, in Fig.~\ref{fig:Ueff} we plot a model potential $U(r)$
for the $^{46}$Mg+$^{46}$Mg reaction. It will be discussed in Sec.\
\ref{s:example}. In this case, $E_C$=16.27 MeV and $\delta$=0.0332,
so that Eq.\ ({\ref{e:Upar}) yields $R_C$=12.53 fm and
$R_{C1}$=12.95 fm. We show the pure Coulomb and parabolic segments
(the solid and dash-dot lines, respectively) separated by the open
dot. The filled dot is the potential maximum. The dotted line is the
pure Coulomb potential extended to $r \to 0$. The thin solid and
dashed vertical lines position, respectively, the maximum and
separation points. Although $\delta$ is formally small, it produces
a noticeable $U(r)$ wing at low $r$.

With the potential (\ref{e:U}) the integral (\ref{e:Phi}) is taken
analytically. At $E < E_{C1}$ we have
\begin{eqnarray}
    \Psi(E) & = & \Psi_r(E)+\Psi_l(E),
\label{e:Phi1} \\
    \Psi_r(E) & = &  4 \sqrt{E_R \over E}\, \left( \arcsin
    \sqrt{x_r} +
    \sqrt{x_r (1-x_r)} \right),
\nonumber \\
    \Psi_l(E) & = &
     -\gamma \, \sqrt{E_R \over E_C}\, { (E_C-E) \over E_C}
\nonumber \\
      && \times
      \left( {\pi \over 2} + \arcsin x_l + x_l \sqrt{1-x_l^2}
     \right),
\nonumber
\end{eqnarray}
where $\gamma=(2+3 \delta)^{3/2} \sqrt{\delta}/(1+\delta)^2$;
$\Psi_r(E)$ and $\Psi_l(E)$ contain the contributions from the
integration regions of $R_{C1} \le r \le r_2$ and $r_1 \le r \le
R_{C1}$, respectively; $x_r=R_{C1}/r_2=R_{C1}E/\alpha=E/E_{C1}$, and
$x_l=\delta \sqrt{\beta E_C/(E_C-E)}$. The term $2 \pi \eta$ exactly
canceled the opposite term which appeared after the integration in
$\Phi(E)$. At $E_{C1}\leq E \leq E_C$ we have
\begin{equation}
   \Psi(E)=2 \pi \eta + \Psi_l(E) =\pi  \left(
    2\,\sqrt{E_R \over E } -  \gamma  \,
   { E_C-E \over
   E_C} \sqrt{E_R \over E_C} \right).
\label{e:Phi2}
\end{equation}

Equations (\ref{e:Phi1}) and (\ref{e:Phi2}) fully determine
$\Psi(E)$ in (\ref{e:below}) in an analytic form. Then
Eqs.~(\ref{e:below}) and (\ref{e:above}) give an analytic, easily
computable model for $S(E)$. It contains four parameters, $S_0$,
$E_C$, $\delta$ and $\xi$; each parameter has simple physical
meaning.

For astrophysical applications, one needs $S(E)$ at subbarrier
energies. In this case, it is natural to present $S(E)$ in the form
\begin{equation}
   S(E)=S(0)\, \exp(g_1E+g_2E^2+\ldots),
\label{e:lowE}
\end{equation}
where $g_1$, $g_2$,\ldots are some expansion coefficients
(determined by low-$r$ behavior of $U(r)$; see below). The main
energy dependence of $S(E)$ at $E<E_C$ is thought to be given by
$\exp(g_{1}E)$. To reduce the energy dependence
one often introduces (e.g., \cite{PWZ1969,Spillane2007}) the
modified $S$-factor
\begin{equation}
     \widetilde{S}(E)=S(E)\exp(-g_{1}E),
\label{e:modifS}
\end{equation}
that is a much less variable function
than $S(E)$; the modified $S$-factor is usually treated as constant.
Our model differs from the traditional approach: instead of
$\widetilde{S}$ we prefer to introduce $S_0$. Both quantities are
nearly constants (at $E \lesssim E_C$) but $S_0$ changes within much
narrower limits  than $\widetilde{S}$ for different reactions
(Sec.~\ref{s:example}).

In our model, we can use Eq.~(\ref{e:below}) and expand the exponent
argument in powers of $E$. Keeping three lowest expansion terms, at
$E$ below $E_C$ we come to Eq.~(\ref{e:lowE}) with
\begin{equation}
   S(0)=\widetilde{S}(0)=S_0\, \exp(g_0),
\label{e:lowE1}
\end{equation}
$g_0$ being the zero-order expansion coefficient (followed by $g_1$,
$g_2$,\ldots). The coefficients can be presented as
$g_i=g_{ir}+g_{il}$ ($i=0,1,2,\ldots$); $g_{ir}$ and $g_{il}$
collect, respectively, the contributions from the right (Coulombic,
$r>R_{C1}$) and left (parabolic, $r \leq R_{C1}$) segments of the
barrier potential. We obtain
\begin{eqnarray}
&&  g_{0r}= 8 \sqrt{ E_R \over E_{C1}}, \quad
  g_{1r}=-{4 \over 3E_{C1}}\, \sqrt{E_R \over E_{C1}},
\nonumber \\
 && g_{2r}=-{1 \over 5E_{C1}^2 } \, \sqrt{ E_R \over E_{C1}};
\label{e:br}\\
&&  g_{0l}= -\gamma \sqrt{E_R \over E_C}\,
  \left( {\pi \over 2} + \arcsin x_{l0}
  + x_{l0} \sqrt{1- x_{l0}^2 }    \right),
\nonumber \\
&&  g_{1l}= { \gamma \over E_C} \sqrt{E_R \over E_C} \,
   \left( {\pi \over 2} + \arcsin x_{l0} \right),
\nonumber \\
&&   g_{2l}= { \gamma \over4 E_C^2} \sqrt{E_R \over E_C} \,
    {x_{l0} \over \sqrt{1-x_{l0}^2}},
\label{e:bl}
\end{eqnarray}
with $x_{l0}=\sqrt{\delta/(2+3 \delta)}$. We can write
$g_0=\sqrt{E_R/E_C}\, \varphi_0(\delta)$,
$g_1=-\sqrt{E_R/E_C^3}\,\varphi_1(\delta)$,
$g_2=\sqrt{E_R/E_C^5}\,\varphi_2(\delta)$,
where $\varphi_i(\delta)$ are functions of
the only one argument $\delta$. For instance, in the limit of
$\delta \ll 1$  the first two functions are
$\varphi_0(\delta)=8 - \pi \sqrt{2 \delta}-2 \delta$
and
$\varphi_1(\delta)={4 \over 3} - \pi \sqrt{2 \delta}- \delta$.

The expansions like (\ref{e:lowE}) have been written long ago (e.g.,
\cite{fh64} and references therein) but only for the sharply
truncated (rectangular -- rct) Coulomb potential,
\begin{equation}
   U_\mathrm{rct}(r)={\alpha \over r}~~~\mathrm{at}~~r\ge R_{C},\quad
   U_\mathrm{rct}(r)=-V_0~~~\mathrm{at}~~r<R_{C}.
\label{e:Usharp}
\end{equation}
This potential is obtained from our potential (\ref{e:U}) in the
limit of $\delta \to 0$, $E_{C1} \to E_C$, and $R_{C1} \to R_C$. In
this case one usually used Eq.~(\ref{e:lowE}) neglecting $g_2$ and
higher-order terms:
\begin{eqnarray}
  &&  S_\mathrm{rct}(E)=S_\mathrm{rct}(0)\,\exp(g_{1\rm rct}E),
\nonumber \\
  &&  g_{1\rm rct}=g_{r1}=- {4 \over 3 \hbar}\,\sqrt{\mu R_{C1}^3 \over 2
    \alpha},
\label{e:Srct}
\end{eqnarray}
with $S_\mathrm{rct}(0)=\widetilde{S}_\mathrm{rct}=S_{0\rm
rct}\,\exp(g_{r0})$. Now $E_{C1}=E_C=\alpha/R_{C1}$; $R_{C1}$ should
be treated as an effective radius of nucleus-nucleus interaction;
the parameters $g_{ir}$ are not well defined because the radius
$R_C=R_{C1}$ of sharp barrier truncation is unphysical. One has
$g_\mathrm{1rct}<0$, that is $S_\mathrm{rct}(E)$ always decreases
with increasing $E$ in the  model (\ref{e:Usharp}). We will show
that it is more reasonable to use more advanced models of $U(r)$
with the broadened peak and $g_1=g_{1r}+g_{1l}$.

\subsection{General properties of model S-factors}
\label{s:features}


The proposed model for $S$-factors is simple and can be analyzed in
general form.

\paragraph{Super-barrier energies.}

The model $S$-factor at $E>E_C$ is given by Eq.~(\ref{e:above}). It
is determined by three constants, $E_C$, $S_0$ and $\xi$, being
independent of $\delta$. This $S$-factor rapidly decreases with the
growth of $E$.

Nevertheless, at $E>E_C$ it is better to use the reaction
cross-section instead of $S(E)$. This cross section reads
$\sigma(E)=S_0\, (E_C\,E)^{-1/2} \, [1+\xi(E-E_C)/E]$; it is a
slowly varying function of energy. Let us stress again that our
Eq.~(\ref{e:above}) cannot be extended to very large $E$, because
the approximation of energy-independent $S_0$ would become
questionable. We expect that the term containing $\xi$
phenomenologically accounts for the contribution of higher partial
waves $\ell>0$ at $E \gtrsim E_C$.

\paragraph{Subbarrier energies.}

\begin{figure*}[t]
\begin{center}
\includegraphics[width=16.0cm,angle=0]{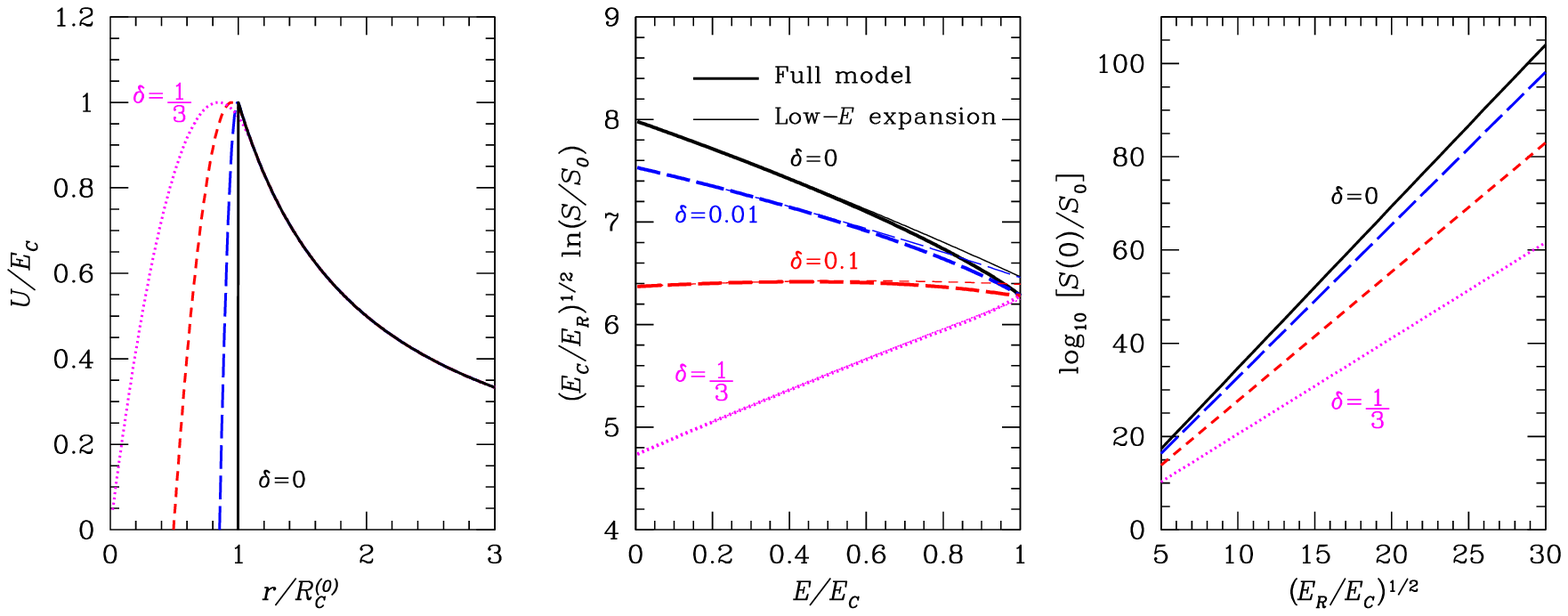}
\caption{(Color online) {\it Left:} Normalized effective potential
$U(r)/E_C$ versus normalized radial coordinate. {\it Mid:}
Normalized  $\ln[S(E)/S_0]$ versus $E/E_C$ [thick lines -- full
model, thin lines -- Eq.~(\ref{e:lowE})]. {\it Right:}
$\log_{10}[S(0)/S_0]$ versus $\sqrt{E_R/E_C}$. Solid, long-dash,
short-dash and dotted lines refer to $\delta$=0, 0.01, 0.1, and 1/3,
respectively.}
\label{fig:tript}
\end{center}
\end{figure*}

In this case, $S(E)$ is determined by $S_0$, $E_C$ and $\delta$,
being independent of $\xi$. Its main features
are illustrated in Fig.~\ref{fig:tript}. The left panel
demonstrates the effective potential $U(r)$. The vertical scale
gives $U(r)/E_C$, which is independent of $E_C$. The horizontal
scale shows $r/R_C^{(0)}$, where $R_C^{(0)}=
\alpha/E_C$ for a given $E_C$.
The solid line gives the sharply truncated Coulomb potential. The
long-dashed, short-dashed, and dotted lines show $U(r)$ for
$\delta=$0.01, 0.1, and $1 \over 3$, respectively. The higher-$r$
wings of $U(r)$ are pure Coulombic, $U(r)=\alpha/r$, independent of
$\delta$. If $E_C$ is fixed but $\delta$ increases, the potential is
broadened at small $r$ and becomes, naturally, less transparent. The
value $\delta={1 \over 3}$ is critical, because nucleus-nucleus
binding at small $r$ disappears [we would have $U(r)\ge 0$ at all
$r\ge 0$]. This case is unrealistic for exothermic nuclear reactions
(with positive $Q$-values) of our study. We show it as a limiting
case.

The middle panel of Fig.~\ref{fig:tript} presents
$\sqrt{E_C/E_R}\,\ln [S(E)/S_0]$ as a function of $E/E_C$ for
subbarrier energies $E\le E_C$ [to visualize the shape of $S(E)$].
As seen from Eqs.~(\ref{e:Phi1}) and (\ref{e:Phi2}), the presented
function depends on the only one parameter, $\delta$. The thick
lines are calculated from Eqs.~(\ref{e:Phi1}) and (\ref{e:Phi2}) at
the same values of $\delta$=0, 0.1, 0.01, and $1 \over 3$ as in the
left panel. Recall that for the sharply truncated Coulomb potential
$S(E)$ {\em increases} as $E \to 0$. For larger $\delta$, the
barrier is less transparent [which can lower $S(E)$ to a great
extent]. Moreover, the $S(E)$ shape becomes different because the
suppression of $S(E)$ at lower $E$ is naturally stronger. In the
range of $0.05 \lesssim \delta \lesssim 0.15$ the $S$-factor turns
into a slowly varying function of $E$, and at higher $\delta$ it
turns into a function which {\em decreases} noticeably with
decreasing $E$, showing a pronounced hindrance of $S(E)$ at low $E$.

The thin lines in the middle panel of Fig.~\ref{fig:tript} present
the same function, as the thick lines, but calculated using the
low-$E$ asymptote (\ref{e:lowE}). This asymptote is seen to be
remarkable accurate for all values of $\delta \leq {1 \over 3}$ and
for all energies $E$ below $E_C$. Actually, the quadratic term
$g_2E^2$ in (\ref{e:lowE}) is small; it is often sufficient to use
the reduced expression
%
$
     S(E)=S_0 \exp(g_0+g_1E).
$
%
The quadratic term modifies the exponent argument maximum by 3\% at
$E=E_C$ and $\delta=0$. Using Eqs.~(\ref{e:br}) and (\ref{e:bl}) one
can show that $g_1$ changes sign at $\delta=0.07745$ (from minus at
smaller $\delta$ to plus at higher). This explains a slow energy
dependence of $S$ at $0.05 \lesssim \delta \lesssim 0.15$. At
$\delta={1 \over 3}$ the function $\ln S(E)$ is almost linear in
$E$, so that the thick and thin dotted lines nearly coincide.

The right panel of Fig.\ \ref{fig:tript} demonstrates $\log_{10}
[S(0)/S_0]$ as a function of $\sqrt{E_R/E_C}$ for the same four
values of $\delta$. For a fixed value of $E_R/E_C$, the $S(0)/S_0$
ratio depends only on $\delta$, in our model. The displayed range of
$\sqrt{E_R/E_C}$ corresponds to the reactions which we consider in
Sec.\ \ref{s:example}. The plot shows the values
$S(0)/S_0=\exp(g_0)$ which determine the zero-energy $S$-factor,
$S(0)$, important for astrophysical applications. For the pure
Coulomb potential extended to $r \to 0$, we would have $S(0)=S_0$.
The potential cutoff at low $r$ drastically enhances the barrier
penetrability, and, hence, $S(0)$ [as well as, generally, low-energy
$S(E)$] over $S_0$, up to $\sim$100 orders of magnitude! The highest
enhancement is provided by the sharply truncated barrier
(\ref{e:Usharp}). If we fix $E_C$ (and thus $E_R/E_C$) and increase
$\delta$, the enhancement can be strongly reduced. Some examples are
given below.

\section{Example: 946  reactions involving C, O, Ne and Mg isotopes}
\label{s:example}
%

\subsection{Calculations}
\label{s:calculations}
%

For example, let us consider a set of $S$-factors, which we have
calculated recently \cite{paper1} for fusion reactions involving
various isotopes of C, O, Ne, and Mg. The calculations include
stable, proton-rich, neutron-rich, and very neutron-rich isotopes.
Such isotopes can appear during nuclear burning in stellar matter,
particularly, in dense matter of white dwarf cores and neutron star
envelopes. The calculations were performed using the S\~ao Paulo
(SP) potential in the context of the barrier penetration model; we
employed the NL3-parametrization of nuclear density distributions of
reactants within the Relativistic Hartree-Bogoliubov (RHB) approach.
The model is based on the standard partial wave decomposition
($\ell=0,1,\ldots$) and considers motion of the nuclei in the
effective potential
\begin{equation}
     U_\mathrm{eff}(r,E)= U_C(r)+U_\mathrm{SP}(r,E)+
     \frac{ \hbar^2 \ell (\ell+1)}{2 \mu r^2},
\label{e:veff}
\end{equation}
where $U_C(r)$, $U_\mathrm{SP}(r,E)$ and the last term are the
Coulomb, nuclear and centrifugal potentials, respectively. At low
energies ($E \lesssim E_C$), the main contribution to $S(E)$ comes
from the $\ell=0$ (s-wave) channel. The calculational scheme is
parameter-free and relatively simple for generating a set of data
for many non-resonant reactions involving different isotopes.

\begin{table}
\caption[]{Fusion reactions $(A_1,Z_1)+(A_2,Z_2)$ under
consideration (after Ref.~\cite{paper1})}
\renewcommand{\arraystretch}{1.2}
\label{tab:reactions}
\begin{center}
\begin{tabular}{c c c c c c }
\hline \hline Reaction & $\quad A_1 \quad $ & $\quad A_2 \quad $ &
~$E_\mathrm{max}$~ & Nr.\ of  \\
type   & even & even &  MeV & cases  \\
\hline
C+C   & 10--24 & 10--24 & 17.9 & 36  \\
C+O   & 10--24 & 12--28 & 17.9 & 72  \\
C+Ne  & 10--24 & 18--40 & 19.9 & 96  \\
C+Mg  & 10--24 & 20--46 & 19.9 & 112  \\
O+O   & 12--28 & 12--28 & 19.9 & 45  \\
O+Ne  & 12--28 & 18--40 & 21.9 & 108  \\
O+Mg  & 12--28 & 18--46 & 21.9 & 126  \\
Ne+Ne & 18--40 & 18--40 & 21.9 & 78  \\
Ne+Mg & 18--40 & 20--46 & 24.9 & 168  \\
Mg+Mg & 20--46 & 20--46 & 29.9 & 105  \\
\hline \hline
\end{tabular}
\end{center}
\end{table}

The reactions in question are summarized in Table
\ref{tab:reactions}. All isotopes studied were even-even nuclei. We
considered 10 reaction types, such as C+C and O+Ne, with the range
of mass numbers for both species given in Table \ref{tab:reactions}.
For each reaction, we computed $S(E)$ on a dense grid of $E$ (with
the energy step of 0.1 MeV) from 2~MeV to a maximum value
$E_\mathrm{max}$ (also given in Table \ref{tab:reactions}) covering
wide energy ranges below and above the Coulomb barrier. The last
column in Table \ref{tab:reactions} presents the number of
considered reactions.

\renewcommand{\arraystretch}{1.2}
\begin{table*}
\caption[]{Fit parameters of $S(E)$ for reactions
$(A_1,Z_1)+(A_2,Z_2)$ under consideration} \label{tab:fit}
\begin{center}
\begin{tabular}{c c c c c c c c c c c c }
\hline \hline Reaction & ~~~$R$~~~~ & ~~$\Delta R_{1a}$~~~ &
~~$\Delta R_{2a}$~~~& ~~$\Delta R_{1b}$~~~ & ~~$\Delta R_{2b}$~~~ &
~~~~~~$\delta$~~~~& ~~~$S_0$~~~& ~~~~$\xi_0$~~~~~& ~~~~~$\xi_1
~~$~~~& ~~Max. &
Rms \\
type   &
fm &
fm  &
fm &
fm  &
fm &
 &
~~MeV b~~ &
   &
   &
dev.   &
dev.   \\
\hline
C+C     &
7.4836  &
0.1759  &
0.1759  &
0.0040  &
0.0040  &
0.0400  &
1.3736  &
3.5499  &
0.2658  &
0.35    &
0.11 \\
C+O      &
7.8671   &
0.1740   &
0.1280   &
--0.0045 &
--0.0310 &
0.0412   &
1.5438   &
5.2576   &
0.2306   &
0.47     &
0.12 \\
C+Ne  &
7.9387 &
0.1720 &
0.1206 &
--0.0171 &
--0.0035 &
0.0400   &
1.9478   &
3.7661   &
0.2328   &
0.62     &
0.16 \\
C+Mg  &
8.0513 &
0.1705 &
0.1014 &
--0.0210 &
--0.0186 &
0.0386 &
2.4327 &
4.0059 &
0.1844 &
0.56  &
0.17  \\
O+O &
8.0641 &
0.1266 &
0.1266 &
--0.0377 &
--0.0377 &
0.0388 &
2.1998 &
6.0147 &
0.1547 &
0.50 &
0.14 \\
O+Ne  &
8.1191 &
0.1257 &
0.1183 &
--0.0461 &
--0.0068 &
0.0371   &
2.9486   &
3.5127   &
0.1702   &
0.65 &
0.19 \\
O+Mg  &
8.2404 &
0.1246 &
0.0994 &
--0.0500 &
--0.0216 &
0.0357 &
3.7433 &
2.8303 &
0.1417 &
0.65 &
0.20 \\
Ne+Ne &
8.1419 &
0.1175 &
0.1175 &
--0.0107 &
--0.0107 &
0.0348 &
4.2215 &
0.1225 &
0.1717 &
1.00 &
0.27 \\
Ne+Mg &
8.2880 &
0.1160 &
0.0987 &
--0.0157 &
--0.0273 &
0.0339 &
5.2525 &
0.2141 &
0.1342 &
0.86 &
0.28  \\
Mg+Mg &
8.4509 &
0.0976 &
0.0976 &
--0.0288 &
--0.0288 &
0.0332 &
5.9785 &
0.5263 &
0.1393 &
0.82 &
0.28 \\
\hline \hline
\end{tabular}
\end{center}
\end{table*}

The results of calculations using the SP model have been compared
previously \cite{yak2006, gas2005,saoPauloTool} with experimental
data (if available) as well as with theoretical calculations
performed using other models such as coupled-channels and fermionic
molecular dynamics ones. As detailed in \cite{saoPauloTool},  the
calculated values of $S(E)$ are uncertain due to nuclear physics
effects -- due to using the SP model with the NL3 nucleon density
distribution. For the reactions involving stable nuclides, typical
uncertainties are expected to be within  a factor of 2, with maximum
up to a factor of 4. For the reactions involving unstable nuclei,
typical uncertainties were roughly estimated to be as large as one
order of magnitude, reaching two orders of magnitude at low energies
for the reactions with very neutron-rich isotopes. These
uncertainties reflect current state of art in our knowledge of
$S(E)$.

\subsection{Fits}
\label{s:Fits}
%

In Ref.~\cite{paper1} we fitted the calculated $S(E)$ by a
9-parameter phenomenological analytic expression. These fits are
accurate (with maximum relative errors less than 10\%) but their use
requires extensive tables (of $9 \times 946=8514$ parameters). Here,
we employ our new fit expressions (Sec.~\ref{s:model}) and show that
the same data can be approximated using only 90 fit parameters.

Let us consider reactions of each type (each line in Table
\ref{tab:reactions}) separately and apply our analytic model
(\ref{e:below}) and (\ref{e:above}) to every reaction. In this way
we determine 4 fit parameters, $S_0$, $E_C$, $\delta$ and $\xi$, for
every reaction. For instance, we have $4 \times 105=420$ parameters
for Mg+Mg reactions. However, we notice that we can put $S_0$ and
$\delta$ constant for all reactions of a given type (for instance,
$S_0= 5.9785$ MeV~b and $\delta=0.0332$ for all Mg+Mg reactions);
this does not increase essentially the fit errors. Such constant
$S_0$ and $\delta$ are given in Table \ref{tab:fit}.

Still, we need to specify two parameters, $E_C$ and $\xi$, for
every reaction. Collecting the values of $E_C$ and $\xi$
for all reactions of
each type, we were able to fit them by analytic expressions
\begin{eqnarray}
      &&E_C =  {\alpha / R_C^{(0)}},
\label{e:ECfit} \\
      && R_C^{(0)}=R+
       \Delta R_1 \, |A_1-A_{10}|+\Delta R_2 \, |A_2-A_{20}|;
\nonumber \\
      && \xi  =  \xi_0+ \xi_1 (A_1+A_2),
\label{e:xifit}
\end{eqnarray}
where $A_{10}=2Z_1$ and $A_{20}=2Z_2$ are mass numbers of most
stable isotopes;
$\Delta R_1=\Delta R_{1a}$ at $A_1 \geq A_{10}$;
$\Delta R_1=\Delta R_{1b}$ at $A_1 < A_{10}$;
$\Delta R_2=\Delta R_{2a}$ at $A_2 \geq A_{20}$;
$\Delta R_2=\Delta R_{2b}$ at $A_2 < A_{20}$.
Thus, we have seven new fit parameters
$R$, $\Delta R_{1a}$, $\Delta R_{2a}$, $\Delta R_{1b}$, $\Delta R_{2b}$;
$\xi_0$ and $\xi_1$
(also given in Table \ref{tab:fit}) for each reaction type, and,
hence, 9 parameters in total. Naturally, we have $\Delta R_2=\Delta
R_1$ for the reactions involving isotopes of the same element (e.g.,
Mg+Mg).

The eleventh column of Table \ref{tab:fit} gives maximum relative
deviation of fitted $S(E)$ from calculated ones for all reactions of
a given type over all energy grid points (e.g., over $105 \times
280= 29400$ points for the Mg+Mg reactions).  We see that the fitted
values of $S(E)$ do not deviate from the calculated ones  by more
than 100\%. Root-mean square (rms) relative deviations, given in the
last column, are a factor of 3--4 lower than the maximum ones. This
fit accuracy is acceptable because it is well within nuclear physics
uncertainties of calculated $S$-factors (Sec.~\ref{s:calculations}).

\subsection{Discussion}
\label{s:discussion}
%

Let us outline the main features of our fits for all 10 reaction
types.

We start with the Mg+Mg reactions (Fig.\ \ref{fig:mgmg}). They are
characterized by the largest strength of Coulomb interaction
(largest product $Z_1 Z_2$ and largest $\alpha$). On the left panel
of Fig.\ \ref{fig:mgmg} we compare calculated and fitted
%
$S(E)$ for six selected Mg+Mg reactions. Similar comparison for
reactions of other types is given below. The solid curves are our
fits; filled dots are calculated $S(E)$. Here and below we plot the
calculated data on a rarefied grid  (with the energy step of 1 MeV)
to simplify the figures. Open dots refer to $E=E_C$ as determined
from Eq.\ (\ref{e:ECfit}). For  each reaction type  we present six
$S(E)$ curves chosen in the following way. The lower curve
corresponds to the reaction with the lightest isotopes
($^{20}$Mg+$^{20}$Mg in Fig.\ \ref{fig:mgmg}). The upper curve is
for the most massive isotopes ($^{46}$Mg+$^{46}$Mg in Fig.\
\ref{fig:mgmg}). The second curve from the bottom is for the most
stable isotopes ($^{24}$Mg+$^{24}$Mg); the third curve is for the
lightest isotope 1 and the most massive isotope 2
($^{20}$Mg+$^{46}$Mg); and two next curves are for somewhat heavier
isotope 1 and lighter isotope 2 ($^{30}$Mg+$^{40}$Mg;
$^{40}$Mg+$^{40}$Mg). The general trend is: the higher the reduced
mass $\mu$ of the reacting nuclei, the larger $S(E)$. We see that
our $S(E)$-model reproduces the data reasonably (uniformly) well for
all Mg+Mg reactions. All these fits are done with the same
$\delta=0.0332$ and $S_0$=5.9785 MeV~b. Fitting the same data with
the model (\ref{e:Usharp}) of rectangular potential
would be much less accurate. This point is also illustrated on the
left panel of Fig.\ \ref{fig:mgmg}. The dotted lines are our best
fits of the same data with the rectangular potential. Such a
potential leads to a faster growth of $S(E)$ at $E \to 0$; it
strongly (up to several orders of magnitude) overestimates
low-energy $S(E)$; fit errors become much higher.

\begin{figure*}[tbh]
\begin{center}
\includegraphics[width=14.0cm,angle=0]{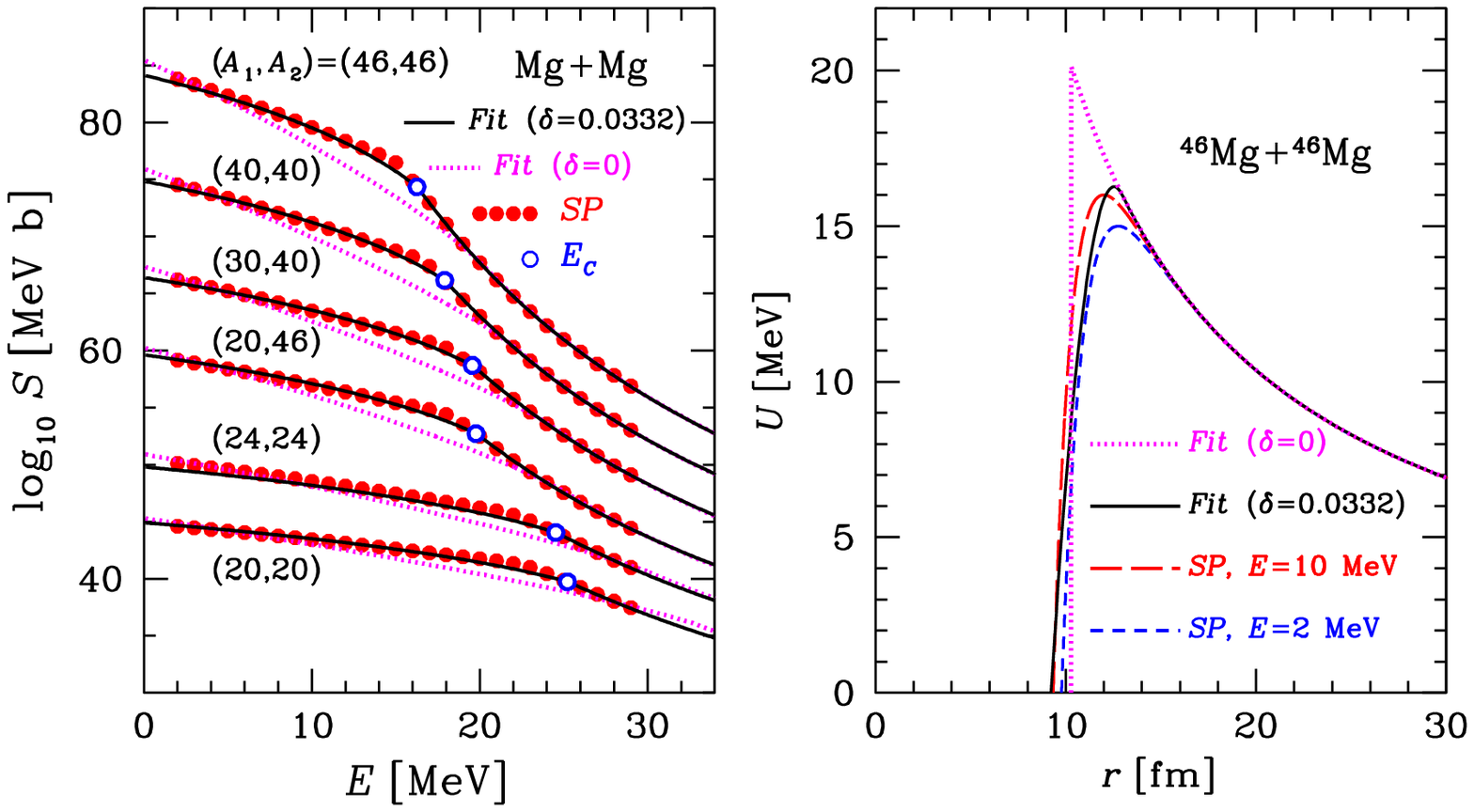}
\caption{(Color online) \textit{Left:} $S$-factors for six Mg+Mg
reactions. Filled dots are the S\~ao Paulo (SP) calculations; solid
lines are our fits (Table \ref{tab:fit}); open dots show the fit
values of $E_C$; dotted lines are best fits of $S(E)$ assuming
$\delta=0$. \textit{Right:} The effective potential $U(r)$ for the
$^{46}$Mg+$^{46}$Mg reaction. The solid line, marked as
\textit{Fit}, is reconstructed from the calculated $S(E)$ using our
model (\ref{e:U}) for $U(r)$; the dotted line is the same but
assuming rectangular potential ($\delta=0$). The long-dashed and
short-dashed lines show the effective potential used in the original
SP calculations of $S(E)$ at $E=$10 and 2 MeV, respectively.}
\label{fig:mgmg}
\end{center}
\end{figure*}

In the right panel of Fig.\ \ref{fig:mgmg} we display the effective
potential $U(r)$ for the $^{46}$Mg+$^{46}$Mg reaction (involving the
most massive Mg isotopes from our collection). The solid line is our
model $U(r)$, reconstructed by fitting calculated values of $S(E)$
with our model. It is given by Eq.\ (\ref{e:U}) and plotted also in
Fig.~\ref{fig:Ueff}, with the fit parameters deduced from our fits
(Table \ref{tab:fit}). The dotted line is our model $U(r)$ for the
best fit with $\delta=0$. The long-dashed and short-dashed lines are
the effective potentials $U_\mathrm{eff}(r,E)$ given by
Eq.~(\ref{e:veff}) and used in original SP calculations. They depend
on $E$ and are plotted for $E$=10 and 2 MeV, respectively. We see
that the reconstructed potential with $\delta=0.0332$ is remarkably
close to the original ones. Therefore, by fitting the available
$S(E)$ data (calculated or experimental ones) with our $S(E)$-model,
one can reconstruct the effective potential $U(r)$. On the other
hand, the model with $\delta=0$ gives the potential (the dotted line
in the right panel of Fig.\ \ref{fig:mgmg}) with unreasonably high
$E_C=20.14$ MeV that is sharply truncated at too large $R_C=10.29$
fm. Naturally, this potential strongly overestimates $S(E)$ at low
$E$.

The value of $S(0)$ for the $^{46}$Mg+$^{46}$Mg reaction is
approximately 83 orders of magnitude higher than our fit value
$S_0$=5.9785 MeV~b. This huge difference is solely attributed to the
definitions of $S(0)$ and $S_0$. While $S_0$ is related to the
penetrability of the real barrier (truncated at small $r$), $S(0)$
is defined through the penetrability of the barrier which remains
pure Coulomb to $r \to 0$. These penetrabilities are drastically
different.

\begin{figure*}[t]
\begin{center}
\includegraphics[width=16.5cm,angle=0]{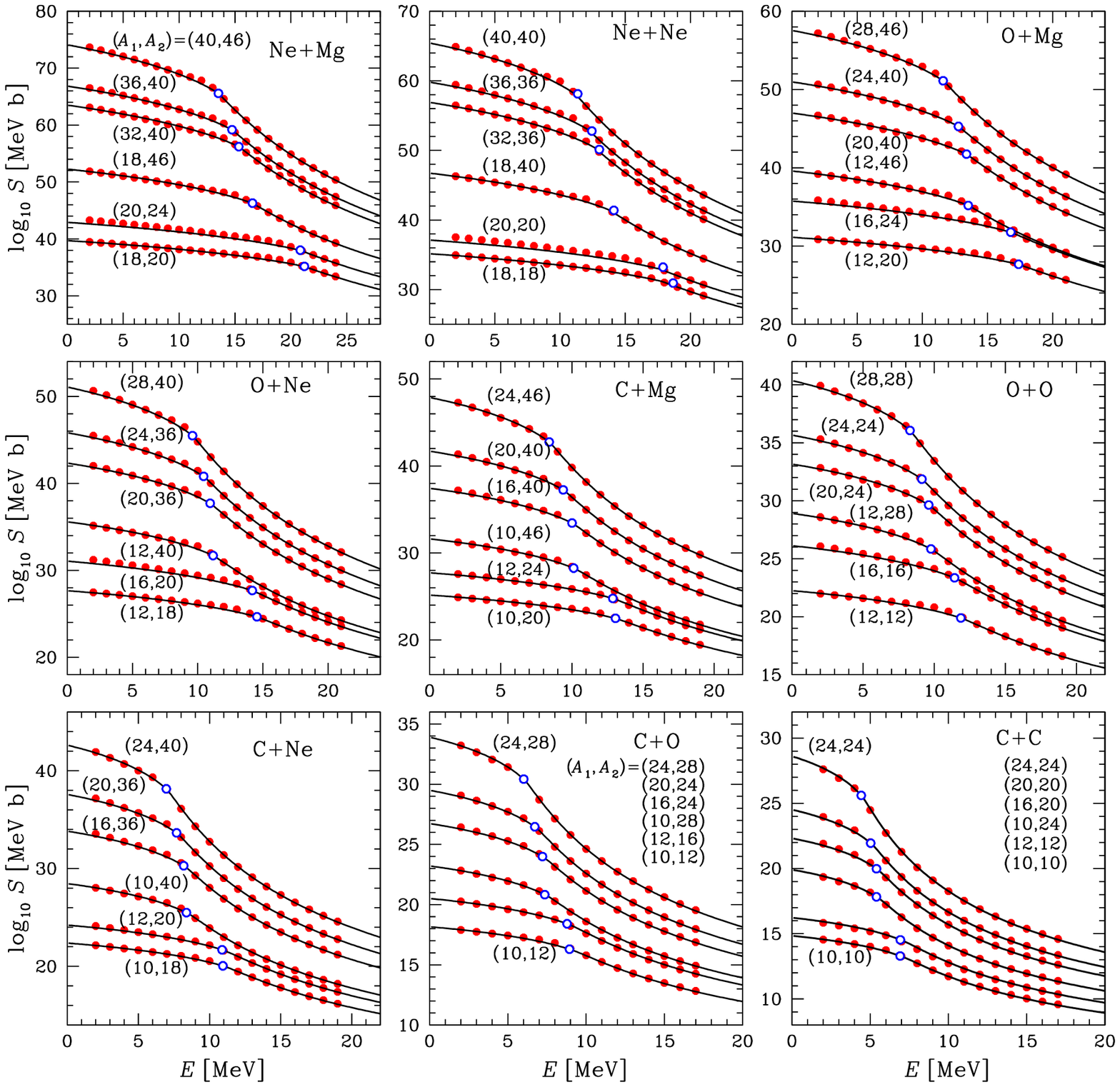}
\caption{(Color online) Each of nine panels shows $S$-factors for
six reactions of one type (as in the left panel of Fig.\
\ref{fig:mgmg}). The reaction types are Ne+Mg, Ne+Ne, O+Mg, O+Ne,
C+Mg, O+O, C+Ne, C+O, and C+C. } \label{fig:eight}
\end{center}
\end{figure*}

Figure \ref{fig:eight} gives nine plots, similar to that on the left
panel of Fig.\ \ref{fig:mgmg}, for the reactions of nine types.
These reactions are ordered (from Ne+Mg, Ne+Ne and O+Mg in the top
row to C+Ne, C+O and C+C in the bottom row) in such a way to have
progressively lower strength of Coulomb nucleus-nucleus interaction
(lower $Z_1 Z_2$). Generally, the fits seem satisfactory and
reasonably uniform.

It is remarkable that the fit parameter $S_0$ takes on the values in
a narrow range from $\approx 1.4$ to $\approx 6$ MeV~b for all
reactions of our study, while the astrophysical $S$-factor varies
over many orders of magnitude. Our approach opens a possibility to
extrapolate the values of $S_0$ to a wider class of reactions
without performing new calculations of $S(E)$. One can also
extrapolate the values of $\delta$, $E_C$ and $\xi$, and obtain thus
$S(E)$-factors for new reactions. Note that we could have fitted all
the data (Table \ref{tab:reactions}) by using one and the same
$\delta \approx 0.04$, and the fit accuracy would be nearly the same
as in our present fits.

Our model $S(E)$ is flexible to describe different $S(E)$ curves. We
believe that the description of $S(E)$ in terms of four well defined
parameters ($S_0$, $\delta$, $E_C$, and $\xi$) is physical and
sufficient for the majority of applications. However, our
interpolation of $E_C$ and $\xi$ by Eqs.~(\ref{e:ECfit}) and
(\ref{e:xifit}) can be regarded only as a reasonably successful
phenomenological fit. We expect that, while doing more accurate
fitting of $S(E)$ for these or other reactions in the future, one
can find better (and physically meaningful) interpolation
expressions for $E_C$ and $\xi$ as functions of $A_1$ and $A_2$, and
consider $\delta$ and $S_0$ as functions of $A_1$ and $A_2$ as well.
One can also improve our fit at $E \gtrsim E_C$ by going beyond the
semi-classical approximation and by replacing the phenomenological
$S(E)$ dependence (\ref{e:above}) with the results of more accurate
consideration. We think that the $S(E)$-factor at $E>E_C$ can be
calculated in a more rigorous form and expressed through $E_C$,
$S_0$ and $\delta$ without introducing an additional parameter like
$\xi$.

Note that for each reaction type we consider some reactions
involving proton-rich nuclei (e.g., $^{10}$C+$^{10}$C) and many
reactions involving neutron-rich nuclei (e.g., $^{24}$C+$^{24}$C).
For each reaction type, we have observed a change in the behavior of
fit parameters on $A_1$ and $A_2$ while crossing the stability line
($A \approx 2Z$). Because we include only a few proton-rich
isotopes, we do not recommend to extrapolate our fits to the region
of proton-rich nuclei (that would require calculations of
$S$-factors for more proton-rich isotopes). We stress that our data
sets include only even-even nuclei. In the future we can
additionally calculate the $S$-factors for reactions involving other
nuclei (even-odd or odd-odd) and approximate them with our model in
a similar fashion. We do not recommend directly extrapolating our
present results (Table \ref{tab:fit}) to these reactions.

\section{$^{12}$C+$^{12}$C reaction}
\label{s:12C}

\begin{figure*}[t]
\begin{center}
\includegraphics[width=14.0cm,angle=0]{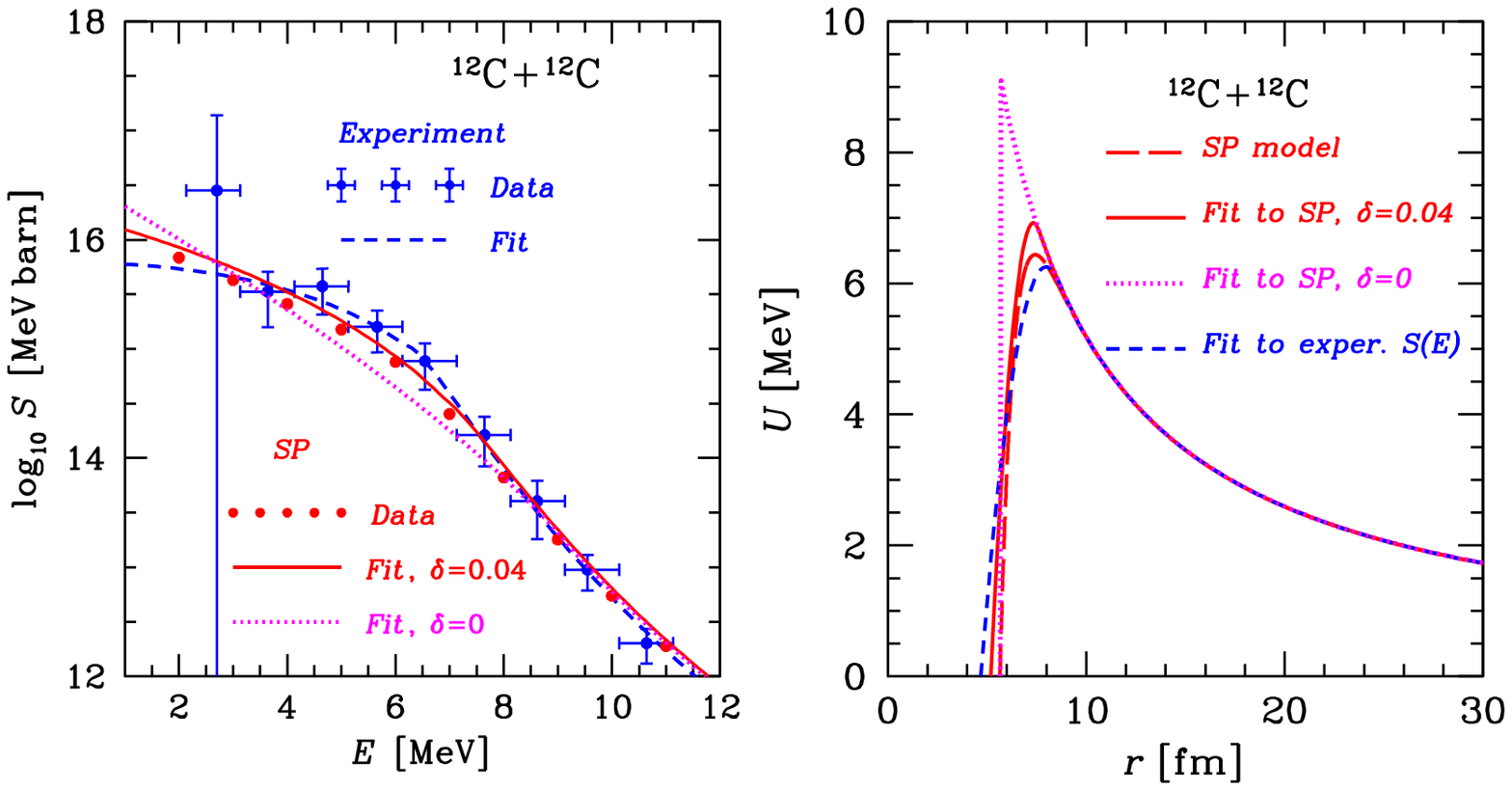}
\caption{(Color online) \textit{Left:} $S$-factors for the
$^{12}$C+$^{12}$C reaction. Crosses show experimental data,
short-dashed line is the best-fit. Filled dots are SP calculations,
solid line is our fit with $\delta=0.04$ (Table \ref{tab:fit});
dotted line is our best fit with $\delta=0$. \textit{Right}: The
effective potential $U(r)$ for the same reaction. Short-dashed line
is reconstructed from the fit to experimental data; long-dashed line
is the original SP potential; solid and dotted lines are the
potentials $U(r)$ reconstructed from the fits to SP data with
$\delta=0.04$ and 0, respectively. See text for details. }
\label{fig:cc}
\end{center}
\end{figure*}

Finally, let us discuss the quality of our $S(E)$ model for the
$^{12}$C+$^{12}$C reaction. The reaction is very important for
astrophysics of compact stars -- for late stellar burning, ignition
of type Ia supernovae and triggering explosive events such as
superbursts (e.g., \cite{hindrance,csb09} and references therein).
Our results are summarized in Fig.~\ref{fig:cc}. On the left panel
we show $S(E)$-data; on the right panel we give corresponding
effective potentials $U(r)$.

The $^{12}$C+$^{12}$C reaction cross section has been measured by
many groups. The cross section contains resonances and the
non-resonant part. We can compare our theoretical model with the
non-resonant contribution. We take experimental $S(E)$ from Refs.\
\cite{PWZ1969,MaS1973,HiC1977,KLR1980,Bec1981,ErB81,das82,Sat82,rosales2003,agu06,Bar06,Spillane2007}.
The data cover the energy range from $\approx 2.1$ MeV to 20 MeV.
The data are nonuniform and not fully consistent, especially at
lower $E$, where experimental $S(E)$ measurements are most difficult
and experimental uncertainties are high. Experimental values of
$S(E)$ seem reasonably accurate at $E \gtrsim 4$ MeV, moderately
uncertain in the $3-4$ MeV range, and rather uncertain at lower $E$.
The presence of low-energy resonances (e.g.,
\cite{Spillane2007,agu06} and references therein) complicates data
analysis. To compare with the SP theoretical calculations, which
neglect the resonances, and to smooth out the effect of experimental
uncertainties we bin the experimental data (with the bin width
$\Delta E=1$ MeV). The binned data are shown by crosses in the left
panel of Fig.\ \ref{fig:cc}. Theoretical SP $S(E)$ values are
presented by filled dots.

The short-dashed line in the left panel of Fig.\ \ref{fig:cc} is our
4-parameter best fit of experimental $S(E)$ factors. In this case we
have $E_C=6.25$ MeV, $\delta=0.0763$, $S_0=0.735$ MeV~b, and $\chi =
13.05$. Reduced chi-square (per one degree of freedom) is 0.41,
which means that the fit is acceptable. The solid line is our fit to
the SP data. It is seen to be in reasonable agreement with the
experimental data. It agrees also with the fit to the experimental
data at those energies at which the data are reliable. The dotted
curve is another fit to the SP data, this time assuming a
rectangular barrier (\ref{e:Usharp}) ($\delta=0$). It is in poor
agreement with the experimental and SP data.

Our 4-parameter fit to SP data gives  (Table \ref{tab:fit})
$E_C=6.93$ MeV, $\delta=0.04$, $S_0=1.37$ MeV~b, $\chi=9.93$. At
subbarrier energies this fit is well described by
Eqs.~(\ref{e:lowE}) and (\ref{e:lowE1}), with the expansion terms
$g_i$ given by (\ref{e:br}) and (\ref{e:bl}). This yields
\begin{equation}
    S_\mathrm{SP}(E)=1.7 \times 10^{16}\,\exp(-0.302E-0.021E^2)
    \quad \mathrm{MeV~b},
\label{e:S_SP}
\end{equation}
where $E$ is in MeV. Note that the contribution $g_{il}$ of the
parabolic segment of the potential barrier $U(r)$ to
$g_i=g_{ir}+g_{il}$ is generally substantial. For instance,
$g_{0}=42.615-5.555=37.060$. This expansion term is absorbed in
$S(0)=1.7 \times 10^{16}$~MeV~b; the value $g_{0l}=-5.555$ reduces
considerably $S(0)$. In the next term $g_1=-1.045+0.743=-0.302$ the
parabolic $U(r)$ part ($g_{1l}=0.743$) is very essential. In the
quadratic term $g_2=-0.023+0.002=-0.021$ it is less important, and
the term itself is rather unimportant in (\ref{e:S_SP}).


A careful analysis of experimental data of different groups has
recently been performed by Aguilera et al.\ \cite{agu06}. The
authors rescaled some data trying to obtain a unified description of
$S(E)$ including resonant structures. Then they took several
theoretical models of non-resonant $S_\mathrm{nr}(E)$ and compared
them with their full unified experimental $S(E)$ data. The
difference $S(E)-S_\mathrm{nr}(E)$ was treated as the resonance
contribution. The best theoretical $S_\mathrm{ns}(E)$ model was
claimed to be that obtained with the KNS (Krappe-Nix-Sierk
\cite{kns79}) barrier potential $U_\mathrm{KNS}(r)$. It gives most
reasonable reproduction of resonant structures after subtracting
non-resonant term.

It is customary (e.g.,
\cite{PWZ1969,clayton83,agu06,Spillane2007,csb09}) to approximate
the non-resonant $S$-factor for the $^{12}$C+$^{12}$C reaction at
subbarrier energies by
\begin{equation}
    S_\mathrm{nr}(E)=\widetilde{S}_a \,\exp(-0.46E),
\label{e:S_app}
\end{equation}
where $\widetilde{S}_a$ is a parameter [analogous to
$\widetilde{S}$, defined by (\ref{e:modifS}), with the specific
value of $g_1=-0.46$ discussed below]. The results of Aguilera et
al.~\cite{agu06} can be approximated in this way using their Fig.~11
[in our notations, that figure gives
$\widetilde{S}(E)=S_\mathrm{nr}(E) \exp(0.46E)$]. Their best
$S_\mathrm{nr}(E)$ (given by the KNS model) in the energy range from
$\approx 3.5$ to $\approx 5.5$ MeV (below the barrier but at those
$E$ where experimental data are reliable) can be approximated by
(\ref{e:S_app}) with $\tilde{S}_a \approx (1.4-1.7) \times 10^{16}$
MeV~b (earlier studies gave $\widetilde{S}_a \approx 3 \times
10^{16}$ MeV~b; e.g., \cite{PWZ1969,cf88}). However, this
approximation becomes inaccurate at lower $E$. We remark, that at
any $E \lesssim 5.5$ MeV, the best (KNS) model of Ref.\ \cite{kns79}
is accurately approximated by ${S}_\mathrm{KNS}(E) \approx 0.83
\times 10^{16} \exp(-0.32E)$ MeV~b, in reasonable agreement with our
approximation (\ref{e:S_SP}) of SP data. Nevertheless, another
acceptable $S_\mathrm{nr}(E)$ model of Ref.\ \cite{agu06}, based on
the proximity-adiabatic (PA) approach \cite{bs81}, is approximated
at the same energies $E \lesssim 5.5$ MeV as ${S}_\mathrm{PA}(E)
\approx 1.6 \times 10^{16} \exp(-0.46E)$ MeV~b, in agreement with
(\ref{e:S_app}).

These different approximations reflect uncertainties in our
knowledge of $S_\mathrm{nr}(E)$. One should be careful in using
(\ref{e:S_app}) for the $^{12}$C+$^{12}$C reaction. The factor $g_1$
in the exponent argument can be different from $g_1=-0.46$ (can be
closer to $-0.3$). The value $g_1=-0.46$ was first introduced by
Patterson et al.\ \cite{PWZ1969}. The authors claimed that it came
from the model of rectangular potential barrier. According to
Eq.~(\ref{e:Srct}), in this model we would have
$g_\mathrm{1rct}=-0.05 \, R_{C}^{3/2}$, where $R_{C}=R_{C1}$ is in
fm. Taking the standard value $R_{C}\approx 2 \times 1.3\, A^{1/3}=
6$ fm we would get $g_\mathrm{1rct}=-0.73$, noticeably different
from $-0.46$. Taking $g_{1rct}=-0.46$ one would have $R_{C}=4.4$ fm
\cite{PWZ1969}, an unrealistically small radius of the potential
well $U(r)$. In fact, Patterson et al.\ \cite{PWZ1969} obtained
$g_1=-0.46$ by fitting a restricted set of experimental $S(E)$
values available by 1969. They discussed possible variations of
$g_1$ but the discussion has been forgotten, whereas their best
value $g_1=-0.46$ is widely cited in the literature.

The right panel of Fig.\ \ref{fig:cc} presents the effective
potentials $U(r)$ corresponding to $S(E)$ models on the left panel.
The short-dashed curve gives $U(r)$ that is calculated from
Eq.~(\ref{e:U}) with the parameters $E_C=6.25$ MeV and
$\delta=0.0763$ inferred from the fit to experimental $S(E)$. It is
our reconstruction of the real potential $U(r)$ (whose details are
still unknown) using our analytic $S(E)$ model. The long-dashed line
is the theoretical SP $U_\mathrm{eff}(r)$ (that is almost
independent of $E$ for the $^{12}$C+$^{12}$C reaction); it was used
to calculate the SP $S(E)$. The solid curve is $U(r)$ reconstructed
from the full ($E_C=6.93$ MeV, $\delta=0.04$) fit to our calculated
SP $S(E)$. The dotted curve is a similar reconstruction but based on
the rectangular $U(r)$ model [Eq.~(\ref{e:Usharp}), $\delta=0$]. The
dotted curve looks unrealistic -- it gives too large $V_C=9.18$ MeV,
indicating once more that the model of rectangular potential is too
crude. Three other potentials are remarkably close even in this,
most difficult $^{12}$C+$^{12}$C case, complicated by pronounced
resonance structures of experimental $S(E)$. This fact confirms that
the SP model is generally a valid tool for studying non-resonant
fusion reactions. Moreover, we see that our analytic $S(E)$ model
can help to reconstruct $U(r)$ from experimental data. Let us add
that the solid, short-dashed and long-dashed $U(r)$ curves are also
close to effective potentials, particularly, to
$U_\mathrm{KNS}(r)$and $U_\mathrm{PA}(r)$, used by Aguliera et al.\
\cite{agu06} to approximate non-resonant contribution to $S(E)$ in
experimental data. As seen from Fig.~8 and Table 3 in \cite{agu06},
the basic parameters ($E_C$ and $R_C$) of $U_\mathrm{KNS}(r)$and
$U_\mathrm{PA}(r)$ are fairly close, but $U_\mathrm{KNS}(r)$ has a
slightly more extended low-$r$ wing which, however, changes
$S_\mathrm{nr}(E)$ behavior at low $E$ (from $g_\mathrm{1KNS}=-0.32$
to $g_\mathrm{1PA}=-0.46$).

Clearly, different segments of $S(E)$ are determined by different
parts of $U(r)$.
The range of $E$ below $E_C$ down to $3-4$ MeV is controlled by
$U(r)$ at $r \gtrsim 6$~fm [not far from the $U(r)$-peak]. In this
case, different theoretical $U(r)$-models give reasonably similar
non-resonant $S(E)$-factors, which generally agree with (accurate)
experimental data. At lower energies, $E \sim 1-3$ MeV, which are
important for $^{12}$C burning in stellar matter in the
thermonuclear regime, experimental values of $S(E)$ are either
uncertain or not available. These values are sensitive to the sharp
low-$r$ wing of the $U(r)$ potential ($r \sim 5-6$ fm) which is not
very well constrained by theory and experiment. However, because the
$1-3$ MeV energy range is close to the range of higher $E$, where
the $S(E)$-factor is well studied, one expects that an extrapolation
from higher $E$ to the $1-3$ MeV range is more or less reliable.
Finally, lowest energies $E \lesssim 1$ MeV are important for
pycnonuclear burning of $^{12}$C in dense stellar matter (e.g.,
\cite{gas2005,yak2006}). In this case, $S(E)$ is controlled by the
very steep low-$r$ slope of $U(r)$ and seems rather uncertain. It
can be affected by slight variations of the $U(r)$ slope (as
discussed above taking KNS and AP models in \cite{agu06} as an
example). Extrapolations to these energies can be inaccurate. The
problem is further complicated by the oblateness of $^{12}$C nuclei
in the ground state. Our analysis is based on the approximation of
spherically symmetric nuclei and spherically symmetric potential
$U(r)$. In the presence of oblateness, $S(E)$-factors depend on
orientations of colliding nuclei. This effect is beyond the scope of
the present paper. Note, however, that according to calculations
\cite{dp10} the oblateness increases low-energy $S(E)$ for the
$^{12}$C+$^{12}$C reaction by a factor of $\sim 1.7$ [which is
within theoretical uncertainties of non-resonant $S(E)$].

We have also compared calculated and fitted $S(E)$-factors with
experimental data for the $^{12}$C+$^{16}$O and $^{16}$O+$^{16}$O
reactions. We have carried out SP calculations based on the RHB
approach outlined in Sec.~\ref{s:calculations}, as well as SP
calculations which employ two-parameter Fermi (2pF) parametrization
of nuclear density distributions of reactants. 2pF calculations
agree with experimental data better than RHB ones (although the
accuracy of both approaches is sufficient for many applications). If
we restricted ourselves to reactions involving stable nuclei, the
2pF parametrization would be more accurate. However, our main goal
was to obtain a uniform set of theoretical $S(E)$ data for a large
collection of nuclei involving unstable ones (to simulate nuclear
burning in neutron stars and white dwarfs). In this case, SP
calculations based on the RHB approach are favorable.

\section{Conclusions}
\label{s:concl}

We have suggested (Sec.~\ref{s:model}) a simple model with
physically meaningful parameters to describe the astrophysical
$S$-factor as a function of center-of-mass energy $E$ of reacting
heavy nuclei for non-resonant fusion reactions. Our main conclusions
are as follows:

\begin{itemize}

\item For any reaction, the model gives $S(E)$ in an analytic form in
terms of four parameters. They are $E_C$, the height of the Coulomb
barrier; $S_0$ that characterizes the strength of the
nucleus-nucleus interaction neglecting Coulomb interaction; $\delta$
that describes the peak broadening of the effective barrier
potential $U(r)$; and $\xi$ to describe the transition from
subbarrier energies to $E \gg E_C$. The model is expected to be
sufficiently accurate  for energies below and above $E_C$ (up to a
few $E_C$).

\item As an example, we have applied our model to describe the
$S$-factors for 946 fusion reactions involving various isotopes of
C, O, Ne, and Mg, from the stability valley to very neutron-rich
nuclei. In Ref.~\cite{paper1} these $S$-factors were calculated
using the SP method and the barrier penetration model. They were
fitted by a phenomenological formula containing 9 fit parameters for
every reaction (8514 parameters in total). With the present analytic
model, we can fit the same data set using 90 fit parameters
(Sec.~\ref{s:Fits}, Table \ref{tab:fit}). The fit accuracy is worse
than in  Ref.~\cite{paper1} but is well within estimated
nuclear-physics uncertainties of calculated $S(E)$
(Sec.~\ref{s:calculations}).

\item We have also compared our model (Sec.~\ref{s:12C}) for the $^{12}$C+$^{12}$C
reaction (that is most important for neutron stars and white dwarfs)
with experimental $S(E)$ data and discussed the problem of
extrapolation of experimental $S(E)$ to low energies of
astrophysical importance.

\item Our analytic $S(E)$-model is easy for implementing into
computer codes, which calculate nuclear reaction rates and simulate
various nuclear burning phenomena in astrophysical environment; it
is not as costly for CPU time as reading large tables can be. The
$S(E)$ dependence for any reaction is determined by the values of
$E_C$, $S_0$, $\delta$ and $\xi$ through Eqs.\ (\ref{e:below}),
(\ref{e:above}) and (\ref{e:Phi1}). As a rule, one needs only
subbarrier $S(E)$ to calculate nuclear reaction rates in stellar
matter. In this case, it is sufficient to use a simplified
expression (\ref{e:lowE}), where the coefficients $g_1$, $g_2$, and
$g_3$ are given by Eqs.\ (\ref{e:br}) and (\ref{e:bl}).

\item The analytic model is practical for describing large uniform
sets of $S(E)$ data (for instance, many reactions involving isotopes
of the same elements). The parameters $E_C$, $S_0$, $\delta$ and
$\xi$ vary slowly from one reaction to another, and are easily
interpolated (by analytic expressions) over large sets; $S_0$ and
$\delta$ can be set constant for many reactions. Analytic
interpolations can be used to extrapolate $S(E)$ to other reactions
of the same type.

\item The functional form of our analytical $S(E)$ is flexible to
describe qualitatively different behaviors of $S(E)$. Particularly,
by varying $\delta$ we can obtain either growth or decrease of
$S(E)$ as $E \to 0$. The decrease is realized at not too small
$\delta$ and may explain the low-energy hindrance of $S(E)$ whose
signature was observed in some reactions (e.g.,
\cite{Jiang07,hindrance} and references therein). The low-energy
behavior of $S(E)$ is indeed very sensitive to the parameter
$\delta$. The model of sharply truncated Coulomb potential
(\ref{e:Usharp}), that is widely used in the astrophysical
literature (e.g., Refs.\ \cite{bbfh57,fh64}), can be inaccurate in
extrapolating calculated or measured $S(E)$ to low $E$
(Secs.~\ref{s:discussion} and \ref{s:12C}).

\item Fitting a given $S(E)$ (computed or measured in laboratory) with
our analytic model can be used to reconstruct the effective
potential $U(r)$ (Figs.\ \ref{fig:mgmg} and \ref{fig:cc}). Of
course, the real potential $U(r)$ can be too complicated to be
exactly described by our model potential (\ref{e:U}). However, we
expect that this potential allows one to reproduce correct $S(E)$
behavior in many cases.

\end{itemize}

There is no doubt that the $U(r)$-peak is not sharp, but broadened.
Roughly speaking, this broadening is twofold. First, the potential
peak becomes smoother. Second, the low-$r$ wing of $U(r)$ becomes
less steep. Clearly, the second effect has stronger impact on the
low-energy $S(E)$ than the first one. In our model, both effects are
described by one and the same parameter $\delta$. We can complicate
the model by introducing new parameters but think that the present
version is good as the first step. Let us add that our model is
useful for reactions between heavy nuclei. Astrophysical $S$-factors
for reactions involving light nuclei contain strong resonances
(e.g., \cite{cf88}) which are not described by our model.

Let us add that nuclear reaction rates in dense stellar matter
(especially, in the cores of white dwarfs and envelopes of neutron
stars \cite{st83}) can be greatly affected by plasma screening of
the Coulomb interaction and by the transition to pycnonuclear
burning regime (where zero-point vibrations of nuclei in a strongly
coupled plasma of ions become important). These plasma physics
effects were described by Salpeter and Van Horn \cite{svh69} (also
see \cite{gas2005,yak2006,cdi07,cd09} and references therein). They
modify the interaction potential $U(r)$ but mainly at sufficiently
large $r$, typically higher than nucleus sizes, while we focus on
the nuclear physics effects which influence $U(r)$ at lower $r$. It
is widely thought that the plasma physics and nuclear physics
effects are distinctly different and can be considered separately.
However, we notice that in very dense and not too hot stellar matter
both effects can become interrelated (and should be considered
together).

We expect that the broadening of the $U(r)$ peak is especially
important for pycnonuclear reactions in the inner crust of accreting
neutron stars in X-ray transients \cite{hz90,hz03,Brown06}. They are
compact binary systems containing a neutron star and a low-mass
companion. Pycnonuclear reactions are thought to be responsible for
deep crustal heating of accreted matter. The heating can power
\cite{Brown98} thermal surface emission of these neutron stars that
is observed in quiescent states of transients (see, e.g., Refs.\
\cite{Brown06,lh07}). Pycnonuclear reactions occur at high densities
and involve very neutron-rich nuclei (e.g., $^{34}$Ne+$^{34}$Ne at
$\rho \approx 2 \times 10^{12}$ g~cm$^{-3}$, according to
Ref.~\cite{hz90}) immersed in a sea of free neutrons (e.g.,
\cite{hpy07}). The $U(r)$-peak should be broadened not only by a
diffusive structure of neutron-rich nuclei (that is taken into
account in the SP calculations) but also by the presence of free
neutrons. The latter effect is unexplored, but it can affect $S(E)$,
nuclear reaction rates, the deep crustal heating and its
observational manifestations (e.g., Ref.\ \cite{ygw06}).

\begin{acknowledgments}
The authors are grateful to Andrey Chugunov for critical remarks.
This work was partly supported by the Joint Institute for Nuclear
Astrophysics (NSF-PHY-0822648), the U.S. Department of Energy under
the grant DE-FG02-07ER41459, the Russian Foundation for Basic
Research (grants 08-02-00837 and 09-02-12080), by the State Program
``Leading Scientific Schools of Russian Federation'' (Grant NSh
2600.2008.2), and by the CompStar Program.
\end{acknowledgments}

\end{document}